\documentclass[preprint,prb]{revtex4}
\usepackage{graphicx}
\usepackage{lscape}
\usepackage{amsmath}
\usepackage{dcolumn}
\usepackage{longtable}
\usepackage{bm}
\newcommand{\CM}{\mathbf{C}}

\newcommand{\SM}{\mathbf{S}}

\newcommand{\BRA}[1]{\langle{} #1 \vert}
\newcommand{\KET}[1]{\vert{} #1 \rangle}
\newcommand{\BRAKET}[1]{\left\langle #1 \right\rangle}
\newcommand{\EPS}[1]{\epsilon_{#1}}
\newcommand{\PAR}[2]{\frac{\partial{#1}}{\partial{#2}}}
\newcommand{\rr}{\mathbf{r}}
\newcommand{\RR}{\mathbf{R}}
\newcommand{\dd}{d}
\newcommand{\db}{\mathbf{d}}
\newcommand{\zz}{\mathbf{\zeta}}

\newcommand{\kk}{\mathbf{k}}

\begin{document}
\linespread{1.3} 
\title{Analytical infrared intensities for periodic systems with 
local basis sets}
\author{Artur F. Izmaylov}
\author{Gustavo E. Scuseria}
\affiliation{Department of Chemistry, Rice University, Houston, Texas 77005, USA}
\date{\today}
\begin{abstract}
We report a method for the efficient evaluation of 
analytic infrared (IR) intensities within
generalized Kohn--Sham density functional theory
using Gaussian orbitals and periodic boundary conditions.
A discretized form of the Berry phase 
is used to evaluate a periodic dipole 
moment and its derivatives with respect to in-phase nuclear 
coordinate displacements. 
Benchmark calculations are presented for one-dimensional chains 
of water molecules and poly(paraphenylenevinylene). 
\end{abstract}
\maketitle
\section{Introduction}
Theoretical prediction of vibrational spectra is a valuable
tool for the characterization of compounds and materials which are 
either not synthesized or difficult to conduct experiments on. 
It requires at least two types of data: 
positions of peaks and their intensities. In the vibrational
spectroscopy of periodic systems, peak positions are associated  
with phonon frequencies. Calculation of phonon frequencies from  
analytical second derivatives have recently become 
available for Kohn--Sham (KS)
density functional theory (DFT) and Hartree--Fock (HF) methods,
with plane-waves (PW)\cite{codes2,Baroni:2001/RMP/515}
and Gaussian type orbitals (GTO).
\cite{Hirata:1998/JMST/121,Izmaylov:2007/JCP/144106}
On the other hand, peak intensities have  
received less attention, even though they are necessary 
for comparison of 
theoretical spectra with their experimental counterparts.
Part of the problem is that only in the 90's was their correct 
theoretical 
expression derived.\cite{King-Smith:1993/PRB/1651,Resta:1994/RMP/899}
Since then, first principle 
studies with KS-DFT and PWs have become common for infrared (IR)
and Raman spectra of solids.\cite{Prosandeev:2005/PRB/214307,Sun:2006/PRB/193101,
Hermet:2007/PRB/220102} 
However, we are aware of 
only one work on IR intensities with GTOs.
Although GTOs have many advantages,\cite{Kudin:2000/PRB/16440,
Izmaylov:2007/JCP/144106} 
especially for orbital dependent functionals, GTO-based techniques
are mostly developed in the quantum chemistry community, 
and therefore,
are underrepresented for periodic systems.
Thus, in this work,
we would like to present a simple scheme for evaluation of 
IR intensities for periodic systems within KS-DFT and HF with
GTOs.

The first order IR intensity 
of a fundamental transition 
exciting the $j$th normal mode is
\begin{equation}\label{eq:IRI}
I_j = \frac{N_A\pi}{3c^2}\left({\left|\PAR{d_x}{Q_j}\right|}^2 +
{\left|\PAR{d_y}{Q_j}\right|}^2 +
{\left|\PAR{d_z}{Q_j}\right|}^2\right),
\end{equation}
where $N_A$ is Avogadro's number, $c$ is the speed of light, 
$d_{x-z}$ are the Cartesian components of the dipole moment, and 
$Q_j$ is the normal mode coordinate.\cite{Book/Wilson:1955}
Since any normal mode is a linear combination
of atomic Cartesian coordinates ($\RR_i$),
IR intensities are calculated from $\partial\db/\partial \RR_i$.
There are two main features arising for these derivatives
under periodic boundary conditions (PBC). First,
one should consider derivatives with respect to in-phase
nuclear coordinate displacements
\begin{equation}
\partial\db/\partial \RR_i =
\sum_{g=0}^{\infty} \partial\db/\partial \RR_{ig},
\end{equation}
where $g$ is the index of an unit cell. Although in-phase vibrations
do not represent the whole phonon spectrum of the periodic system,
only these vibrations are IR active.\cite{Book/Decius:1977}
Second, under PBC,
the dipole moment per unit cell 
cannot be straightforwardly expressed as a matrix 
element of the position operator. 
\cite{King-Smith:1993/PRB/1651,Resta:1994/RMP/899,Resta:1998/PRL/1800,
Zak:2000/PRL/1138}
According to the modern theory of polarization,
\cite{King-Smith:1993/PRB/1651,Resta:1994/RMP/899}  
the dipole moment per unit cell is a geometric quantum 
phase or a Berry phase.\cite{Berry:1984/PRSA/45} 
A non-trivial quantum phase usually 
appears in cases when the system under study is coupled to the
rest of the universe through some external parameter $\zz$.
\cite{Resta:2000/JPC/R107} 
The Berry phase $\gamma$ can then be written as 
\begin{equation}\label{eq:bph}
\gamma = {\rm Im} \oint_{\mathcal{C}} d\zz 
\BRAKET{\psi(\zz)\Big{|}\PAR{\psi(\zz)}{\zz}},
\end{equation} 
where $\psi(\zz)$ is the wavefunction and $\mathcal{C}$ is a
closed contour. In the case of 
periodic dipole moment, 
the parameter $\zz$ is represented by a reciprocal vector $\kk$ running 
over the Brillouin zone. Therefore, within single-determinant methods,
Eq.~\ref{eq:bph} necessitates evaluating $\partial/\partial\kk$ 
derivatives of 
the Slater determinant. Since the latter is obtained in 
the self-consistent field (SCF) procedure which usually adds
to the solution arbitrary complex phase prefactors depending on the
$\kk$ value, it is not trivial to differentiate the SCF solution 
with respect to $\kk$. Two main approaches 
to this problem have been proposed. The first method involves a 
discrete 
representation of $\partial/\partial\kk$ 
derivatives through matrix elements
at neighboring $\kk$ points. This representation is 
invariant with respect to phase arbitrariness which could 
arise from the SCF procedure. 
\cite{King-Smith:1993/PRB/1651,Resta:1994/RMP/899} 
The second approach introduces
phase factors which cancel the SCF arbitrary 
phases and allow one to evaluate $\partial/\partial\kk$ 
by employing regular
finite differences techniques. 
\cite{Bishop:2001/JCP/7633,Kirtman:2000/JCP/1294,
Springborg:2007/JCP/104107} 
Here, we start from the periodic dipole expression 
arising in the first approach.
We consider the first approach as a simpler and 
computationally more robust alternative, since it does not require  
introduction of complex phase functions and band resolution
techniques.
Therefore, it can be seen as a black-box technique that 
has only the number of $\kk$ points in the Brillouin zone
sampling as an input.
In addition, this method has been tested 
with plane waves before,
\cite{Sai:2002/PRB/104108,Nunes:2001/PRB/155107} and 
its robustness was demonstrated not only for the linear 
response regime but for finite electric fields as well.
Yet another approach to the periodic dipole derivatives 
is possible in the case of PWs, where these quantities 
require only dipole matrix elements between occupied and 
virtual Bloch functions $\BRA{\phi_{nk}}\rr\KET{\phi_{ak}}$.
These ``transition'' dipole elements can be transformed 
into the following  well-defined expression
\cite{Baroni:1986/PRB/7017,Hybertsen:1987/PRB/5585}
\begin{equation}
\BRA{\phi_{nk}}\rr\KET{\phi_{ak}} =
\frac{\BRA{\phi_{nk}}[H_{\rm SCF},\rr]
\KET{\phi_{ak}}}{\EPS{a}(\kk)-\EPS{n}(\kk)},
\end{equation}
where $H_{\rm SCF}$ is the crystal SCF Hamiltonian,  
$\EPS{a}(\kk)$ and $\EPS{n}(\kk)$ are orbital energies of 
empty and occupied bands. However, in the case of GTOs 
this approach could be less beneficial since Pulay's
type of terms arise. 
To the best of our knowledge, the only implementation 
of IR intensities with GTOs was reported by Jacquemin and coworkers
\cite{Jacquemin:2003/JCP/3956} using the second approach.  
Thus, the focus of the current work is in adapting the 
discretized 
Berry phase approach for evaluation of IR intensities with GTOs.

This work can be also 
seen as complementary to our previous work on zone
centered phonon frequencies.\cite{Izmaylov:2007/JCP/144106} 
One should keep in mind that 
IR intensities are always obtained with vibrational 
frequencies, 
and therefore, any algorithm for IR intensities can be seen
as an efficient one if its execution time is negligible with respect to
that for vibrational frequencies.  

\section{Theory}
\subsection{Berry phase approach to periodic dipole moment}

We first consider the formulation of periodic dipole moment through the 
Berry phase approach in localized basis sets proposed by 
Kudin and coworkers.\cite{Kudin:2007/JCP/234101}
We restrict our consideration to closed-shell
one-dimensional periodic 
systems aligned along the $z$ axes. 
This makes our description simpler without 
losing essential details. Throughout this work
we will be using Bloch orbitals 
\begin{equation}\label{eq:bo}
\phi_{nk} (\rr)= \frac{1}{\sqrt{N_c}}\sum_{g=0}^{N_c}\sum_{\mu=1}^{M}
\mu_g(\rr) C_{\mu n} (k) e^{igka}
\end{equation}
expanded over Gaussian atomic orbitals (AO) $\mu_g(\rr)$.
Here, small Greek letters are for AO indices with
subscripts denoting the unit cell number.
In Eq.~(\ref{eq:bo}), $N_c$ is the total number of unit cells in the system,
$C_{\mu n}(k)$ are crystal orbital (CO) coefficients
from the SCF problem under PBC,\cite{Kudin:2000/PRB/16440} 
$a$ is the length of the translational vector, and $M$ is the 
total number of basis functions per unit cell.  
The cell-periodic part of a Bloch orbital corresponding to the 
$n$th band
\begin{equation}
u_{nk}(\rr) = e^{-ikz} \phi_{nk} (\rr)
\end{equation}
can be expanded as 
\begin{equation}\label{eq:u}
u_{nk}(\rr) = \frac{1}{\sqrt{N_c}}\sum_{g=0}^{N_c}\sum_{\mu=1}^{M}
\mu_g(\rr) C_{\mu n} (k) e^{ik(ag-z)}.
\end{equation}
According to the Berry phase approach,
\cite{King-Smith:1993/PRB/1651,Resta:1994/RMP/899} 
the $z$ component of the dipole moment per unit cell is
\begin{equation}
\dd = -\mathrm{Im}\sum_{n}^{\rm occ.}\int_{\rm BZ} dk 
\int d\rr~ u_{nk}^{*}(\rr)\PAR{u_{nk}(\rr)}{k}.
\end{equation}
Here, we avoid introducing an additional $z$ subscript, 
since unless otherwise stated we always will consider
the longitudinal part of the dipole moment. 
Substituting $u_{nk}(\rr)$ in the dipole expression 
by Eq.~(\ref{eq:u}) we obtain the following representation 
for the periodic dipole moment
\begin{eqnarray}
\dd &=& \dd_1 + \dd_2, \\
\dd_1 &=& \sum_{n}^{\rm occ.}\sum_{\mu,\nu}^{M}\int_{\rm BZ} 
dk~ C_{\mu n}^{*}(k) C_{\nu n}(k) \\
\nonumber
&&\times\sum_{g}e^{ikga}
(Z_{\mu\nu}^{0g}-gaS_{\mu\nu}^{0g}), \\ \label{eq:d2} 
\dd_2 &=& -\mathrm{Im}\sum_{n}^{\rm occ.}\sum_{\mu,\nu}^{M}
\int_{\rm BZ} dk~ C_{\mu n}^{*}(k) \PAR{C_{\nu n}(k)}{k}\\
\nonumber
&&\times\sum_{g}e^{ikga}S_{\mu\nu}^{0g},
\end{eqnarray}
where 
\begin{eqnarray}
Z_{\mu\nu}^{0g} &=& \int d\rr~ \mu_0(\rr) z \nu_g(\rr),
\end{eqnarray}
and 
\begin{eqnarray}
S_{\mu\nu}^{0g} &=& \int d\rr~ \mu_0(\rr) \nu_g(\rr), 
\end{eqnarray}
are dipole and overlap matrices in the AO basis.
Calculation of the $\dd_1$ term is more efficient to
perform in the AO basis 
\begin{eqnarray}
\dd_1 = \sum_{\mu\nu,g} P_{\mu\nu}^{0g}
(Z_{\mu\nu}^{0g}-gaS_{\mu\nu}^{0g}),
\end{eqnarray}
where $P_{\mu\nu}^{0g}$ is the electron density matrix.
The matrix $(Z_{\mu\nu}^{0g}-gaS_{\mu\nu}^{0g})$ can be further split
into Hermitian 
\begin{eqnarray}
\widetilde{Z}_{\mu\nu}^{0g} &=& Z_{\mu\nu}^{0g}-\frac{ga}{2}S_{\mu\nu}^{0g}
\end{eqnarray}
and anti-Hermitian 
\begin{eqnarray}
\bar{Z}_{\mu\nu}^{0g} &=& -\frac{ga}{2}S_{\mu\nu}^{0g}
\end{eqnarray}
parts. Both matrices $\widetilde{Z}_{\mu\nu}^{0g}$ and
$\bar{Z}_{\mu\nu}^{0g}$ are real, and we refer 
to Hermiticity of their reciprocal space counterparts
$\widetilde{Z}_{\mu\nu}(k)$ and $\bar{Z}_{\mu\nu}(k)$.
Since we contract both parts with the Hermitian matrix 
$P_{\mu\nu}^{0g}$, the anti-Hermitian part can be safely omitted.
This leads us to the following working expression
\begin{eqnarray}\label{eq:d1ao}
\dd_1 = \sum_{\mu\nu,g} P_{\mu\nu}^{0g}\widetilde{Z}_{\mu\nu}^{0g}.
\end{eqnarray}

The $\dd_2$ term requires special care because CO 
coefficients usually contain arbitrary $k$-dependent 
phase factors from a diagonalization procedure. Therefore, 
one cannot apply regular numerical differentiation schemes
directly to CO coefficients. To cope with this difficulty
we use a discretization approach\cite{Kudin:2007/JCP/234101} 
which starts from rewriting 
Eq.~(\ref{eq:d2}) in the form 
\begin{equation}\label{eq:tr}
\dd_2 = -\mathrm{Im}\int_{\rm BZ} dk~\mathrm{tr}\left[
C^{\dagger}(k)S(k)\PAR{C(k)}{k}\right],
\end{equation}
where only the occupied orbital part of $C(k)$ is engaged.
For further discussion we introduce the following
$O$-by-$O$ matrix 
\begin{equation}
\Sigma(k,k') = C^{\dagger}(k)S(k)C(k'),
\end{equation}
where $O$ is the number of occupied bands.
Treating $k'$ as a variable and $k$ as a parameter we can write 
\begin{equation}\label{eq:1}
C^{\dagger}(k)S(k)\PAR{C(k)}{k} = \PAR{}{k'}\Sigma(k,k')\Big{|}_{k=k'}.
\end{equation}
Keeping in mind the orthogonality relation $\Sigma(k,k)=1$ we rewrite
Eq.~\ref{eq:1} as
\begin{equation}
\PAR{}{k'}\Sigma(k,k')\Big{|}_{k=k'} = 
\PAR{}{k'}{\rm ln~}\Sigma(k,k')\Big{|}_{k=k'}. 
\end{equation}
The trace operation commutes with the 
differentiation
\begin{equation}\label{eq:trcom}
{\rm tr}\left[\PAR{}{k'}{\rm ln~}
\Sigma(k,k')\Big{|}_{k=k'}\right] = 
\PAR{}{k'}{\rm tr}\left[{\rm ln~}
\Sigma(k,k')\right]\Big{|}_{k=k'}.
\end{equation}
Applying the well-known matrix 
relation ${\rm tr~ln}(A) = {\rm ln~det}(A)$
to the $\Sigma(k,k')$ matrix we obtain
\begin{equation}\label{eq:trdet}
\PAR{}{k'}{\rm tr}\left[{\rm ln~}
\Sigma(k,k')\right]\Big{|}_{k=k'} = 
\PAR{}{k'}{\rm ln}\left[{\rm det~}
\Sigma(k,k')\right]\Big{|}_{k=k'}.
\end{equation}
Since ${\rm ln~det~}\Sigma(k,k) = 0$,
using simple rectangular discretization for 
the $k'$ derivatives and the Brillouin zone integration
we arrive to the following discretized form
\cite{Kudin:2007/JCP/234101} for $\dd_2$:
\begin{eqnarray}\label{eq:dd2} 
\dd_2 &=& -\mathrm{Im}\sum_{j=1}^{N_k}\mathrm{ln~}
\mathrm{det~}\Sigma(k_j,k_{j+1}) \\
&=&-\mathrm{Im}\sum_{j=1}^{N_k}\mathrm{ln~}
\mathrm{det~}C^{\dagger}(k_j)S(k_j)C(k_{j+1}). 
\end{eqnarray}
By introducing arbitrary phase factors $e^{i\theta(k_j)}$ in front 
of CO coefficients, one can easily see that they cancel each other 
after the summation over the Brillouin zone. Please note that one
cannot evaluate Eq.~(\ref{eq:tr}) by using a perturbation 
(linear response) expression to build $C(k+\Delta k)$ from
$C(k)$. As was pointed out in Ref.~\onlinecite{Resta:1994/RMP/899},
a perturbation theory implicitly uses the so called 
``parallel-transport'' gauge which necessitates phase equality 
between $C(k)$ and $C(k+\Delta k)$, and as a consequence
produces zero Berry phase. 

Similar to the molecular case, the
transverse components of the periodic dipole moment
can be written as
\begin{eqnarray}
d_q = \sum_{\mu\nu,g}P_{\mu\nu}^{0g}Q_{\mu\nu}^{0g},
\end{eqnarray}
where $q=x$ or $y$ and 
\begin{eqnarray}
Q_{\mu\nu}^{0g} = \int d\rr \mu_0(\rr) q \nu_g(\rr).
\end{eqnarray}
Therefore, their differentiation is very similar 
to that of the $\dd_1$ part of the longitudinal dipole moment,
and it will not be considered further.

\subsection{Dipole derivatives}

Differentiation of Eq.~(\ref{eq:d1ao})
with respect to nuclear coordinate 
displacements gives 
\begin{eqnarray}\label{eq:d1R}
\PAR{\dd_1}{\RR_i} &=& \sum_{\mu\nu,g}
\PAR{P_{\mu\nu}^{0g}}{\RR_i}\widetilde{Z}_{\mu\nu}^{0g}
+ P_{\mu\nu}^{0g}\PAR{\widetilde{Z}_{\mu\nu}^{0g}}{\RR_i}. 
\end{eqnarray}
Both terms in Eq.~(\ref{eq:d1R}) are calculated with standard 
techniques: the coupled perturbed SCF (CPSCF) procedure for 
the density derivatives,
\cite{Izmaylov:2006/JCP/224105,Izmaylov:2007/JCP/144106}
and a regular integral derivatives evaluation 
technique.\cite{Gill:1991/IJQC/745} 
The CPSCF procedure evaluates the same density derivatives as 
in the case of vibrational frequency calculations;
therefore computational overhead of the $\dd_1$ part  
is negligible, since it only involves calculation of
one-electron terms.

In order to obtain the $\dd_2$ derivatives we adapt the algebraic 
approach discussed in Ref.~\onlinecite{Sai:2002/PRB/104108}. 
The main step can be summarized as application of the 
following algebraic identity 
\begin{eqnarray}\label{eq:d1}
\PAR{}{\RR}\{{\rm ln~det}[\Sigma(\RR)]\}
= {\rm tr}\left[\Sigma^{-1}(\RR)\PAR{}{\RR}\Sigma(\RR)\right],
\end{eqnarray}
which is an analog of Eqs.~\ref{eq:trcom} and \ref{eq:trdet}.
This was proven from a different point of view in 
Ref.~\onlinecite{Sai:2002/PRB/104108}.
The differentiation of $\Sigma(k_j,k_{j+1})$ gives rise to two
types of terms in the $\dd_2$ derivatives
\begin{eqnarray}
\PAR{\dd_2}{\RR_i} = 
\dd_{2}^{(1)}\left(\PAR{\SM}{\RR_i}\right)
+ \dd_{2}^{(2)}\left(\PAR{\CM}{\RR_i}\right),
\end{eqnarray}
which correspond to 
Pulay's type
of forces from overlap integral derivatives
\begin{eqnarray}
\dd_{2}^{(1)} &=& -\mathrm{Im~}\sum_{j=1}^{N_k}\sum_{m,n}^{\rm occ.}
C_{\mu n}^{*}(k_j)\PAR{S_{\mu\nu}(k_j)}{\RR_i}
C_{\nu m}(k_{j+1})\\ \nonumber
&&\times\Sigma_{mn}^{-1}(k_{j},k_{j+1}) \\
\label{eq:d21}
&=& -\mathrm{Im~}\sum_{j=1}^{N_k}\sum_{m,n}^{\rm occ.}
\PAR{S_{n m}(k_j,k_{j+1})}{\RR_i} \\ \nonumber
&&\times\Sigma_{mn}^{-1}(k_{j},k_{j+1}),
\end{eqnarray}
and response of CO coefficients
\begin{eqnarray}
\dd_{2}^{(2)} &=& -\mathrm{Im~}\sum_{j=1}^{N_k}\sum_{m,n}^{\rm occ.}
\Sigma_{mn}^{-1}(k_{j},k_{j+1}) \\ \nonumber
&\times&\Big{[}\PAR{C_{\mu n}^{*}(k_j)}{\RR_i}
S_{\mu\nu}(k_j) C_{\nu m}(k_{j+1}) \\ \nonumber
&+& C_{\mu n}^{*}(k_j)S_{\mu\nu}(k_j)\PAR{C_{\nu m}(k_{j+1})}
{\RR_i}\Big{]}.
\end{eqnarray}
To complete the construction of the dipole moment derivatives we 
will express CO derivatives via the response matrix $U$
obtained in the CPSCF procedure\cite{Hirata:1998/JMST/121}
\begin{eqnarray}\label{eq:linr}
\PAR{C_{\mu n}^{*}(k_j)}{\RR_i} = \sum_{a}^{\rm vir.}
U_{na}^{*}(k_j)C_{\mu a}^{*}(k_j).
\end{eqnarray}
Thus we arrive to 
\begin{eqnarray}\label{eq:d22}
\dd_{2}^{(2)} &=& -\mathrm{Im~}\sum_{j=1}^{N_k}\sum_{m,n}^{\rm occ.}
\sum_{a}^{\rm vir.}\Sigma_{mn}^{-1}(k_{j},k_{j+1}) \\ \nonumber
&&\times\Big{[}U_{na}^{*}(k_j)\Sigma_{am}(k_{j},k_{j+1}) \\ \nonumber
&&+\Sigma_{na}(k_{j},k_{j+1})U_{am}(k_{j+1})\Big{]}.
\end{eqnarray}
Since we do not evaluate $\RR_i$ dependent phases, the application of
the linear response Eq.~(\ref{eq:linr}) is acceptable. 

The computational complexity of the presented scheme 
is negligible with respect to that of vibrational 
frequency calculations. For the $\dd_2$ part we have several matrix
multiplications of $M$-by-$M$ matrices and one
$O$-by-$O$ matrix inversion for each $\kk$ point. 

Two- and three-dimensional generalizations of the presented method
can be done in the same way as for  
plane-wave implementations; therefore, we refer the interested reader 
to Ref.~\onlinecite{Sai:2002/PRB/104108} for details.  

\section{Benchmark calculations}

\subsection{Algorithmic tests}

\begin{figure}[p!]
\centerline{\includegraphics[angle=0,width=12cm]{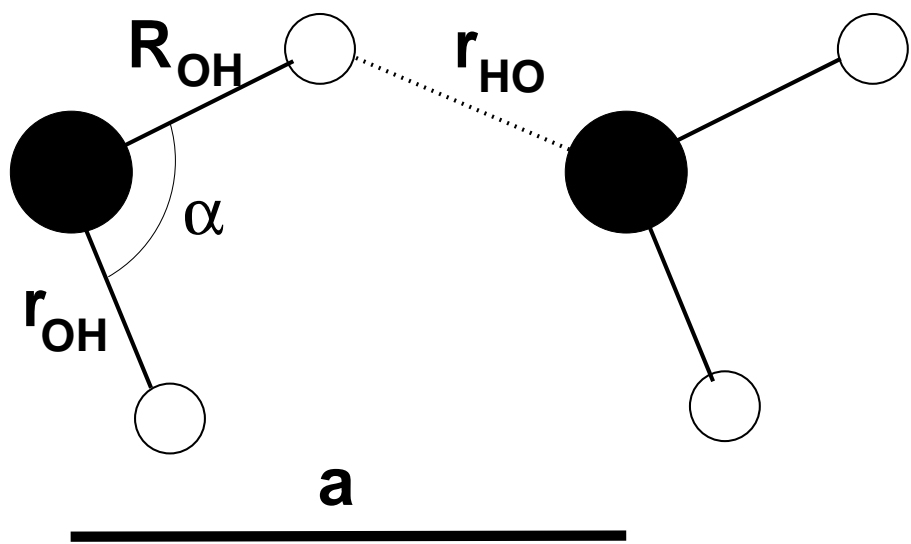}}
\caption{\label{fig:h2o}Schematic structure of the one-dimensional water 
chain. Optimized parameter values are listed in Table \ref{tab:geom}.} 
\end{figure}

To validate and assess the present formalism, which has
been implemented in the development version of the 
\textsc{gaussian} program,\cite{GDV-short} we have
tested it on a 
model one-dimensional chain of water molecules
(see Fig.~\ref{fig:h2o} and Table \ref{tab:geom}) 
with the geometry optimized within
the HF and PBE methods.\cite{gdv:setup} 
In order to describe hydrogen-bond interactions properly, 
we employ the 6-311++G** basis set in most of 
our tests.\cite{Book/Hadzi:1997}

\begin{center}
\begin{table}
\caption{\label{tab:geom} Optimized geometry (\AA~ and degrees) of 
a one-dimensional water chain used to benchmark the accuracy 
of our analytic method with different basis sets. 
See Fig.~\ref{fig:h2o} for the definition of the geometrical 
parameters.}
\begin{ruledtabular}
\begin{tabular}{c c c c c}
& \multicolumn{2}{c}{HF} & \multicolumn{2}{c}{PBE}\\
\cline{2-3}\cline{4-5}
& 6-31G & 6-311++G** & 6-31G & 6-311++G** \\
\hline
$R_{OH}$ & 0.9625 & 0.9477 & 1.0213 & 0.9843 \\
$r_{OH}$ & 0.9492 & 0.9410 & 0.9818 & 0.9692 \\
$r_{HO}$ & 1.9084 & 2.1349 & 1.6367 & 1.9158 \\
$a$ & 2.7459 & 2.9547 & 2.5709 & 2.8048 \\
$\alpha$ & 108.58 & 104.31 & 105.77 & 102.76 \\
\end{tabular}
\end{ruledtabular}
\end{table}
\end{center}

In Table~\ref{tab:olig} we compare dipole
derivatives obtained in periodic calculations with those from 
oligomeric estimations. 
Given that the 6-311++G** basis set is too large
to perform vibrational frequency calculations for long oligomeric
chains, we have used the 6-31G basis set for this comparison. 
The oligomeric results are calculated by the difference 
scheme which is less prone to edge effects
\begin{equation}\label{eq:olig}
\PAR{\mathbf{d}}{\RR_i} = \left(\PAR{\mathbf{d}}
{\RR_i}\right)^{(N)}
-\left(\PAR{\mathbf{d}}{\RR_i}\right)^{(N-1)},
\end{equation}
where 
\begin{equation}
\left(\PAR{\mathbf{d}}{\RR_i}\right)^{(N)} 
= \sum_{m=1}^{N} \PAR{\mathbf{d}}{\RR_{i}^{(m)}},
\end{equation}
here, the indices $i$ and $m$ enumerate atoms within a molecule and 
molecules within an oligomer, respectively.
Numerical periodic results are obtained by 
calculating finite differences between unit cell dipole moments 
in different geometrical configurations. As expected, oligomeric
calculations need more molecules to converge in the periodic 
direction than in the others. The same trend can be seen in 
analytic periodic calculations with respect to the 
number of $\kk$ points.
The numerical PBC calculation is less sensitive to the number
of $\kk$ points, however, they have a fixed error due to the finite
difference scheme of differentiation. This suggests that the periodic
dipole moment converges faster than 
its derivatives with the number of $\kk$ points. 

\begin{center}
\begin{table}
\caption{\label{tab:olig}
Comparison of dipole derivatives with respect to Cartesian coordinates 
of the oxygen atom (in a.u.)
for periodic and oligomeric calculations of the water chain with 
HF
(see Fig.~\ref{fig:h2o}).}
\begin{ruledtabular}
\begin{tabular}{c c c c c}
Method & $\partial d_z /\partial O_z $& $\partial d_z /\partial O_y $ & $\partial d_y /\partial O_z $ & $\partial d_y /\partial O_y $ \\
\hline
\multicolumn{5}{c}{Basis 6-31G}\\
\multicolumn{5}{l}{Oligomeric $n/n-1$}\\
32/31 & -1.02563 & -0.08673 & -0.12885 & -0.51942 \\
52/51 & -1.02580 & -0.08675 & -0.12885 & -0.51942\\
82/81 & -1.02585 & -0.08676 & -0.12885 & -0.51942 \\
102/101 & -1.02587 & -0.08676 & -0.12885 & -0.51942 \\
\multicolumn{5}{l}{Numerical (PBC)}\\
$N_k=$ 1 000 & -1.02611 & -0.08664 & -0.12872 & -0.51931 \\
$N_k=$ 10 000 & -1.02611 & -0.08664 & -0.12872 & -0.51931 \\
\multicolumn{5}{l}{Analytic (PBC)}\\
$N_k=$ 1 000 & -1.02581 & -0.08670 & -0.12885 & -0.51942 \\
$N_k=$ 10 000 & -1.02590 & -0.08676 & -0.12885 & -0.51942 \\
\multicolumn{5}{c}{Basis 6-311++G**}\\
\multicolumn{5}{l}{Numerical (PBC)}\\
$N_k=$ 1 000 & -0.83322 & -0.09828 & -0.08772 & -0.54018 \\
$N_k=$ 10 000 & -0.83322 & -0.09826 & -0.08772 & -0.54018 \\
\multicolumn{5}{l}{Analytic (PBC)}\\
$N_k=$ 1 000 & -0.79599 & -0.10754 & -0.08496 & -0.53669 \\
$N_k=$ 10 000 &-0.79627 & -0.10774 & -0.08496 & -0.53669 \\
\end{tabular}
\end{ruledtabular}
\end{table}
\end{center}

As a matter of practical interest we illustrate the dependence of
the five normal mode frequencies and their 
IR intensities on the number of 
$\kk$ points involved in the discretization of the Brillouin zone
with HF and DFT (see Table~\ref{tab:kdep}).
As a representative DFT method we used the non-empirical
generalized gradient approximation (GGA) functional of
Perdew-Burke-Ernzerhof\cite{Perdew:1996/PRL/3865} (PBE).
These model calculations suggest that 100 - 1000 $\kk$ points
are enough to obtain IR intensities converged within 1\%.
A large difference in sensitivity of vibrational frequencies
and IR intensities to $N_k$ should not be considered as
a serious issue, since the overhead from enlarging $N_k$
is negligible in our vibrational frequency evaluation algorithm.
\cite{Izmaylov:2007/JCP/144106}
We suppose that the large difference in magnitudes of
HF and PBE IR intensities is related to the tendency of
pure DFT functionals to yield a more metallic description than
the HF method.

\begin{center}
\begin{table}
\caption{\label{tab:kdep}
Convergence of vibrational frequencies (cm$^{-1}$) and 
IR intensities (km/mol) with the number of $\kk$
points ($N_k$) for the five normal modes of the water chain 
in HF and PBE  
(see Fig.~\ref{fig:h2o}). Band gaps are denoted by $E_g$.}
\begin{ruledtabular}
\begin{tabular}{c c c c c c c c c c c}
 &\multicolumn{2}{c}{1}&\multicolumn{2}{c}{2}&
\multicolumn{2}{c}{3}&\multicolumn{2}{c}{4}&\multicolumn{2}{c}{5}\\
\cline{2-3}\cline{4-5}\cline{6-7}\cline{8-9}\cline{10-11}
$N_k$ & Freq. & Int. & Freq. & Int. & Freq. & Int. & Freq. & Int. &
 Freq. & Int.\\
\hline
\multicolumn{11}{c}{\quad HF/6-311++G**, $E_g=14.4$ eV}\\
30  & 399.0 &  92.0  & 585.8 &  271.7  & 1785.3 &  146.2  & 4069.0 &  208.5  & 4211.9 &  79.2 \\
50  & 399.0 &  92.8  & 585.8 &  271.7  & 1785.3 &  146.6  & 4069.0 &  210.8  & 4211.9 &  79.7 \\
100  & 399.0 &  93.3  & 585.8 &  271.7  & 1785.3 &  147.0  & 4069.0 &  212.6  & 4211.9 &  80.1 \\
1000  & 399.0 &  93.7  & 585.8 &  271.7  & 1785.3 &  147.4  & 4069.0 &  214.1  & 4211.9 &  80.4 \\
10 000  & 399.0 &  93.8  & 585.8 &  271.7  & 1785.3 &  147.5  & 4069.0 &  214.2  & 4211.9 &  80.5 \\
\multicolumn{11}{c}{\quad PBE/6-311++G**, $E_g=5.3$ eV}\\
30  & 503.6 &  75.8  & 544.6 &  192.7  & 1630.9 &  157.9  & 3489.0 &  426.6  & 3789.1 &  34.8 \\
50  & 503.6 &  76.7  & 544.6 &  192.7  & 1630.9 &  157.2  & 3489.0 &  446.5  & 3789.1 &  35.0 \\
100  & 503.6 &  77.3  & 544.6 &  192.7  & 1630.9 &  156.9  & 3489.0 &  461.4  & 3789.1 &  35.0 \\
1000  & 503.6 &  77.8  & 544.6 &  192.7  & 1630.9 &  156.6  & 3489.0 &  474.6  & 3789.1 &  35.0 \\
10 000  & 503.6 &  77.9  & 544.6 &  192.7  & 1630.9 &  156.6  & 3489.0 &  475.9  & 3789.1 &  35.0 \\
\end{tabular}
\end{ruledtabular}
\end{table}
\end{center}

\subsection{Poly(paraphenylenevinylene)}

\begin{figure}[p!]
\centerline{\includegraphics[angle=0,width=17cm]{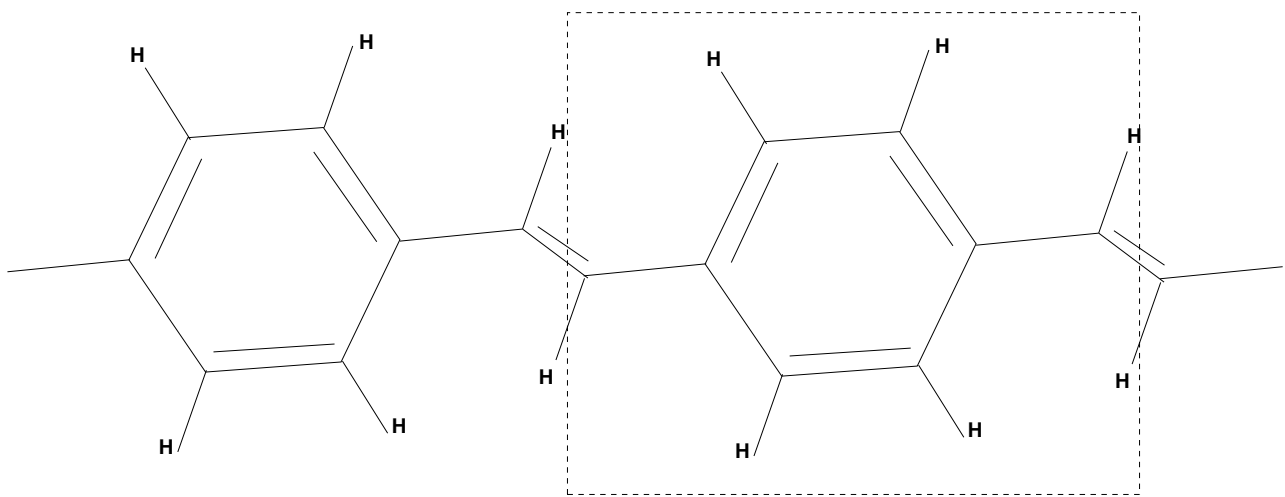}}
\caption{\label{fig:ppv}Poly(paraphenylenevinylene) (PPV).}
\end{figure}

In order to illustrate the prediction capabilities of our approach,
we chose the experimentally and theoretically well-studied  
one-dimensional system 
of poly(paraphenylenevinylene) (PPV, Fig.~\ref{fig:ppv}).
\cite{Burroughes:1990/N/539,
Tian:1991/JCP/3191,Rakovic:1990/PRB/10744,
Kudin:2000/PRB/16440} 
Such an interest is motivated by conductivity and 
luminescent nonlinear properties 
which make this system a promising candidate as a material for 
light-emitting diodes.\cite{Burroughes:1990/N/539}
We would like to point out that previous theoretical work from our group
on PPV was done with some limitations which are overcome
in the current study. In Ref.~\onlinecite{Kudin:2000/PRB/16440},
the IR intensities took into account only the changes in the non-periodic part
of the dipole moment ($\dd_1$).
This caused an underestimation of IR intensities for 
those vibrations that modify the longitudinal dipole moment of PPV.

\begin{figure}[p!]
\centerline{\includegraphics[angle=0,width=15cm]{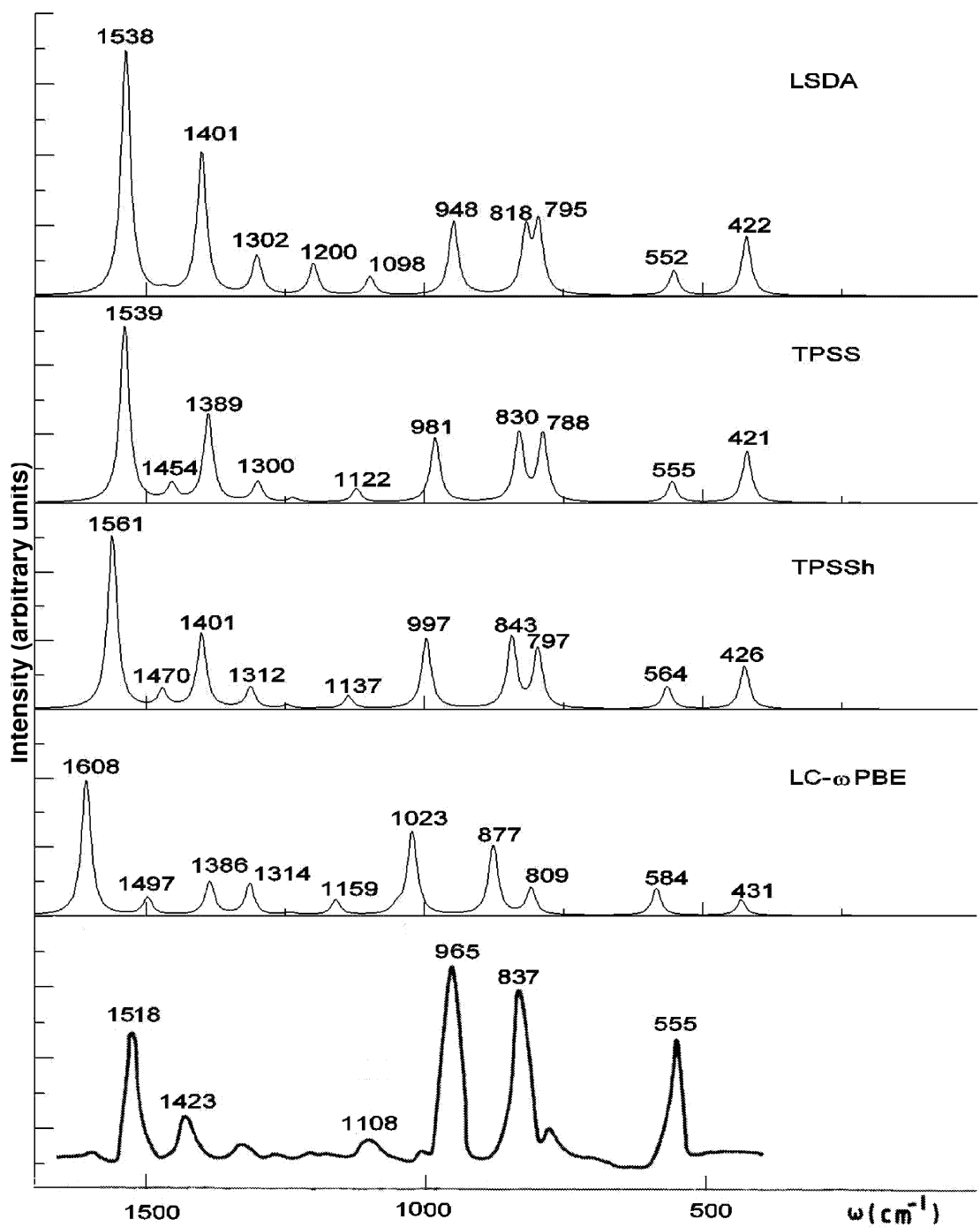}}
\caption{\label{fig:irspec}IR spectra of PPV. The experimental spectrum 
(bottom graph) is based on data given in 
Ref.~\onlinecite{Rakovic:1990/PRB/10744}.
The calculated spectra were obtained by using a Lorentzian 
broadening with a 10 cm$^{-1}$ width.}
\end{figure}

As mentioned in the Introduction,
IR intensities usually receive less attention and their values are 
presented only graphically in arbitrary units.
\cite{Tian:1991/JCP/3191,Rakovic:1990/PRB/10744} 
Therefore, in Table \ref{tab:ppv} 
we compare IR-active calculated vibrational frequencies with their
experimental counterparts. Note that in the case of IR intensities
only a qualitative comparison with experimental data can be done
(see Fig.~\ref{fig:irspec}).
In our calculations we used the following functionals:
the local spin density approximation\cite{Vosko:1980/CJP/1200} (LSDA),
the meta-GGA of Tao-Perdew-Staroverov-Scuseria\cite{Tao:2003/PRL/146401} (TPSS),
TPSS hybrid\cite{Tao:2003/PRL/146401} (TPSSh), and long-range
corrected hybrid PBE\cite{Vydrov:2006/JCP/234109} (LC-$\omega$PBE). 
Our choice was motivated by a previous study of the performance 
of the TPSS functional in molecules and solids.
\cite{Staroverov:2003/JCP/12129,Izmaylov:2007/JCP/144106} 
We also included the LC-$\omega$PBE
functional because we expect some overestimation of IR intensities 
by regular functionals due to  the well-known problem 
with electric field response properties in extended systems.
\cite{Champagne:1998/JCP/10489,Jacquemin:2007/JCP/144105}
According to Table \ref{tab:ppv} and Fig.~\ref{fig:irspec}, 
all functionals perform 
quite adequately, although they generally underestimate 
lower frequencies and overestimate higher ones. Mean absolute 
errors with respect to experimentally observable frequencies 
indicate that the pure DFT methods (LSDA and TPSS) describe 
vibrational frequencies better than the hybrid functionals 
(TPSSh and LC-$\omega$PBE).
Calculated values of IR intensities qualitatively follow
the right trends in most of the cases. It is remarkable  
that even though LC-$\omega$PBE does not perform very well 
for vibrational frequencies, it 
predicts IR intensities which are in a better agreement 
with experiment than are those from the regular functionals. 
This complies with the idea that IR intensities with 
regular functionals suffer mostly from the wrong asymptotic behavior
of the exchange potential. 
It is possible that the correct $1/r$ asymptotic behavior
of the LC-$\omega$PBE exchange potential
\cite{Vydrov:2006/JCP/234109} is responsible for the 
improvements of IR intensities in elongated systems.

\begin{center}
\begin{table}
\caption{\label{tab:ppv}
PPV IR active vibrational frequencies (cm$^{-1}$) and
their intensities (km/mol) calculated with various methods
and the 6-31G** basis set. In all calculations $N_k=1000$ was used.}
\begin{ruledtabular}
\begin{tabular}{c c c c c c c c c c}
 &\multicolumn{2}{c}{LSDA}&\multicolumn{2}{c}{TPSS}&
\multicolumn{2}{c}{TPSSh}&\multicolumn{2}{c}{LC-$\omega$PBE} & 
Expt.\footnotemark[1]\\
\cline{2-3}\cline{4-5}\cline{6-7}\cline{8-9}\cline{10-10}
Mode\footnotemark[2] & Freq. & Int. & Freq. & Int. & Freq. & Int. & Freq. 
& Int. & Freq. \\
\hline
$A_u$ & 224 & 0.2 & 225 & 0.2 & 230 & 0.2 & 241 & 0.2 &   \\
$B_u$ & 422 & 26.5 & 421 & 23.6 & 426 & 19.8 & 431 & 7.1 & 429 \\
$A_u$ & 552 & 11.3 & 555 & 9.7 & 564 & 10.4 & 584 & 12.2 & 
555(s)\footnotemark[3] \\
$B_u$ & 795 & 30.8 & 788 & 31.0 & 797 & 27.2 & 809 & 12.4 & 785 \\
$A_u$ & 818 & 27.9 & 830 & 31.4 & 843 & 32.9 & 877 & 32.1 & 837(s) \\
$A_u$ & 911 & 0.4 & 941 & 0.2 & 959 & 0.2 & 1009 & 1.0 &   \\
$A_u$ & 948 & 33.0 & 981 & 29.9 & 997 & 32.4 & 1023 & 37.5 & 965(s) \\
$B_u$ & 1000 & 4$\times$10$^{-2}$& 1011 & 3$\times$10$^{-2}$ & 1025 & 2$\times$10$^{-2}$ & 1050 & 3.2 & 1013 \\
$B_u$ & 1098 & 8.5 & 1122 & 6.3 & 1137 & 5.9 & 1159 & 7.0 & 1108 \\
$B_u$ & 1200 & 14.0 & 1236 & 2.0 & 1247 & 1.7 & 1240 & 1.1 & 1211  \\
$B_u$ & 1302 & 17.3 & 1300 & 9.5 & 1312 & 9.9 & 1314 & 14.3 & 1271 \\
$B_u$ & 1401 & 63.5 & 1389 & 40.2 & 1401 & 34.4 & 1386 & 15.5 & 1339 \\
$B_u$ & 1467 & 1.5 & 1454 & 7.8 & 1470 & 8.1 & 1497 & 8.1 & 1423(s) \\
$B_u$ & 1538 & 109.6 & 1539 & 81.0 & 1561 & 79.3 & 1608 & 61.8 & 1518(s) \\
$B_u$ & 3061 & 135.4 & 3121 & 114.3 & 3156 & 104.3 & 3220 & 62.7 &   \\
$B_u$ & 3090 & 60.9 & 3136 & 84.9 & 3171 & 77.5 & 3230 & 38.3 &   \\
$B_u$ & 3109 & 20.2 & 3162 & 49.7 & 3197 & 45.1 & 3256 & 21.1 &   \\
\hline
MAE\footnotemark[4]& 21 &  & 17 & & 27 &  & 43 &  &  \\
\end{tabular}
\end{ruledtabular}
\footnotetext[1]{References~\onlinecite{Tian:1991/JCP/3191} and
\onlinecite{Rakovic:1990/PRB/10744}.}
\footnotetext[2]{The normal modes are classified according to the
crystallographic point group C$_{2h}$.}
\footnotetext[3]{``(s)'' stands for strong banks.}
\footnotetext[4]{Mean absolute error with respect to experimentally observed 
frequencies.}
\end{table}
\end{center}

\section{Final remarks}

We have presented a simple route to evaluating IR intensities
in solids within the HF and DFT frameworks with localized 
basis sets. As in the molecular case, to evaluate 
IR intensities one needs to 
differentiate the dipole moment with respect to nuclear 
coordinate displacements. However, the periodic 
dipole moment is not a straightforward generalization 
of its molecular counterpart but rather can be seen 
as a geometric quantum phase.   
Thus, we have used the discretized Berry phase expression for the
dipole moment per unit cell developed in the modern theory of 
polarization\cite{King-Smith:1993/PRB/1651,Resta:1994/RMP/899}
and elaborated for localized basis sets
by Kudin and coworkers.\cite{Kudin:2007/JCP/234101} 
We have differentiated the discretized dipole moment expression 
with respect to 
in-phase nuclei displacements and have demonstrated validity 
of our technique on the model 
one-dimensional chain of water molecules.
The evaluation of IR intensities introduces only a negligible overhead 
in the characterization of vibrational frequencies, since 
CPU timings for dipole derivatives 
[Eqs.~(\ref{eq:d1}), (\ref{eq:d21}), and (\ref{eq:d22})],
in all studied cases constitute less than 4 \% of total CPU times.
The PPV study  
with different functionals 
reveals that although the calculated vibrational frequency adequately 
reproduce those from experiment, the corresponding IR intensities 
do not always follow qualitatively correct trends. Application
of the LC-$\omega$PBE functional corrects IR intensities 
presumably due to 
the right $1/r$ asymptotic behavior of the LC-$\omega$PBE 
exchange potential 
but worsens frequencies. 
We hope that our scheme 
and the results reported here will stimulate  
further functional development in the future.  
  
\begin{acknowledgments}
A.F.I. would like to thank K. N. Kudin, T. M. Henderson, and
O. Hod for valuable suggestions.
This work was supported by the Department of Energy under Grant
No. DE-FG02-04ER15523 and by the Welch Foundation.
\end{acknowledgments}

\begin{thebibliography}{40}
\expandafter\ifx\csname natexlab\endcsname\relax\def\natexlab#1{#1}\fi
\expandafter\ifx\csname bibnamefont\endcsname\relax
  \def\bibnamefont#1{#1}\fi
\expandafter\ifx\csname bibfnamefont\endcsname\relax
  \def\bibfnamefont#1{#1}\fi
\expandafter\ifx\csname citenamefont\endcsname\relax
  \def\citenamefont#1{#1}\fi
\expandafter\ifx\csname url\endcsname\relax
  \def\url#1{\texttt{#1}}\fi
\expandafter\ifx\csname urlprefix\endcsname\relax\def\urlprefix{URL }\fi
\providecommand{\bibinfo}[2]{#2}
\providecommand{\eprint}[2][]{\url{#2}}

\bibitem[{\citenamefont{Baroni et~al.}(2001)\citenamefont{Baroni, de~Gironcoli,
  Corso, and Giannozzi}}]{Baroni:2001/RMP/515}
\bibinfo{author}{\bibfnamefont{S.}~\bibnamefont{Baroni}},
  \bibinfo{author}{\bibfnamefont{S.}~\bibnamefont{de~Gironcoli}},
  \bibinfo{author}{\bibfnamefont{A.~D.} \bibnamefont{Corso}}, \bibnamefont{and}
  \bibinfo{author}{\bibfnamefont{P.}~\bibnamefont{Giannozzi}},
  \bibinfo{journal}{Rev. Mod. Phys.} \textbf{\bibinfo{volume}{73}},
  \bibinfo{pages}{515} (\bibinfo{year}{2001}).

\bibitem[{cod()}]{codes2}
\bibinfo{note}{See for example, http://www.pwscf.org and
  http://www.abinit.org.}

\bibitem[{\citenamefont{Hirata and Iwata}(1998)}]{Hirata:1998/JMST/121}
\bibinfo{author}{\bibfnamefont{S.}~\bibnamefont{Hirata}} \bibnamefont{and}
  \bibinfo{author}{\bibfnamefont{S.}~\bibnamefont{Iwata}}, \bibinfo{journal}{J.
  Mol. Struct. (Theochem)} \textbf{\bibinfo{volume}{451}}, \bibinfo{pages}{121}
  (\bibinfo{year}{1998}).

\bibitem[{\citenamefont{Izmaylov and
  Scuseria}(2007)}]{Izmaylov:2007/JCP/144106}
\bibinfo{author}{\bibfnamefont{A.~F.} \bibnamefont{Izmaylov}} \bibnamefont{and}
  \bibinfo{author}{\bibfnamefont{G.~E.} \bibnamefont{Scuseria}},
  \bibinfo{journal}{J. Chem. Phys.} \textbf{\bibinfo{volume}{127}},
  \bibinfo{pages}{144106} (\bibinfo{year}{2007}).

\bibitem[{\citenamefont{King-Smith and
  Vanderbilt}(1993)}]{King-Smith:1993/PRB/1651}
\bibinfo{author}{\bibfnamefont{R.~D.} \bibnamefont{King-Smith}}
  \bibnamefont{and}
  \bibinfo{author}{\bibfnamefont{D.}~\bibnamefont{Vanderbilt}},
  \bibinfo{journal}{Phys. Rev. B} \textbf{\bibinfo{volume}{47}},
  \bibinfo{pages}{1651} (\bibinfo{year}{1993}).

\bibitem[{\citenamefont{Resta}(1994)}]{Resta:1994/RMP/899}
\bibinfo{author}{\bibfnamefont{R.}~\bibnamefont{Resta}}, \bibinfo{journal}{Rev.
  Mod. Phys.} \textbf{\bibinfo{volume}{66}}, \bibinfo{pages}{899}
  (\bibinfo{year}{1994}).

\bibitem[{\citenamefont{Prosandeev et~al.}(2005)\citenamefont{Prosandeev,
  Waghmare, Levin, and Maslar}}]{Prosandeev:2005/PRB/214307}
\bibinfo{author}{\bibfnamefont{S.~A.} \bibnamefont{Prosandeev}},
  \bibinfo{author}{\bibfnamefont{U.}~\bibnamefont{Waghmare}},
  \bibinfo{author}{\bibfnamefont{I.}~\bibnamefont{Levin}}, \bibnamefont{and}
  \bibinfo{author}{\bibfnamefont{J.}~\bibnamefont{Maslar}},
  \bibinfo{journal}{Phys. Rev. B} \textbf{\bibinfo{volume}{71}},
  \bibinfo{eid}{214307} (\bibinfo{year}{2005}).

\bibitem[{\citenamefont{Sun et~al.}(2006)\citenamefont{Sun, Zhou, Chen, Fan,
  Wang, Guo, He, and Tian}}]{Sun:2006/PRB/193101}
\bibinfo{author}{\bibfnamefont{J.}~\bibnamefont{Sun}},
  \bibinfo{author}{\bibfnamefont{X.-F.} \bibnamefont{Zhou}},
  \bibinfo{author}{\bibfnamefont{J.}~\bibnamefont{Chen}},
  \bibinfo{author}{\bibfnamefont{Y.-X.} \bibnamefont{Fan}},
  \bibinfo{author}{\bibfnamefont{H.-T.} \bibnamefont{Wang}},
  \bibinfo{author}{\bibfnamefont{X.}~\bibnamefont{Guo}},
  \bibinfo{author}{\bibfnamefont{J.}~\bibnamefont{He}}, \bibnamefont{and}
  \bibinfo{author}{\bibfnamefont{Y.}~\bibnamefont{Tian}},
  \bibinfo{journal}{Phys. Rev. B} \textbf{\bibinfo{volume}{74}},
  \bibinfo{eid}{193101} (\bibinfo{year}{2006}).

\bibitem[{\citenamefont{Hermet et~al.}(2007)\citenamefont{Hermet, Goffinet,
  Kreisel, and Ghosez}}]{Hermet:2007/PRB/220102}
\bibinfo{author}{\bibfnamefont{P.}~\bibnamefont{Hermet}},
  \bibinfo{author}{\bibfnamefont{M.}~\bibnamefont{Goffinet}},
  \bibinfo{author}{\bibfnamefont{J.}~\bibnamefont{Kreisel}}, \bibnamefont{and}
  \bibinfo{author}{\bibfnamefont{P.}~\bibnamefont{Ghosez}},
  \bibinfo{journal}{Phys. Rev. B} \textbf{\bibinfo{volume}{75}},
  \bibinfo{eid}{220102(R)} (\bibinfo{year}{2007}).

\bibitem[{\citenamefont{Kudin and Scuseria}(2000)}]{Kudin:2000/PRB/16440}
\bibinfo{author}{\bibfnamefont{K.~N.} \bibnamefont{Kudin}} \bibnamefont{and}
  \bibinfo{author}{\bibfnamefont{G.~E.} \bibnamefont{Scuseria}},
  \bibinfo{journal}{Phys. Rev. B} \textbf{\bibinfo{volume}{61}},
  \bibinfo{pages}{16440} (\bibinfo{year}{2000}).

\bibitem[{\citenamefont{Wilson et~al.}(1955)\citenamefont{Wilson, Decius, and
  Cross}}]{Book/Wilson:1955}
\bibinfo{author}{\bibfnamefont{E.~B.} \bibnamefont{Wilson}},
  \bibinfo{author}{\bibfnamefont{J.~C.} \bibnamefont{Decius}},
  \bibnamefont{and} \bibinfo{author}{\bibfnamefont{P.~C.} \bibnamefont{Cross}},
  \emph{\bibinfo{title}{Molecular Vibrations}} (\bibinfo{publisher}{McGraw-Hill
  Inc.}, \bibinfo{address}{New York}, \bibinfo{year}{1955}).

\bibitem[{\citenamefont{Decius and Hexter}(1977)}]{Book/Decius:1977}
\bibinfo{author}{\bibfnamefont{J.~C.} \bibnamefont{Decius}} \bibnamefont{and}
  \bibinfo{author}{\bibfnamefont{R.~M.} \bibnamefont{Hexter}},
  \emph{\bibinfo{title}{Molecular Vibrations in Crystals}}
  (\bibinfo{publisher}{McGraw-Hill Inc.}, \bibinfo{address}{New York},
  \bibinfo{year}{1977}).

\bibitem[{\citenamefont{Resta}(1998)}]{Resta:1998/PRL/1800}
\bibinfo{author}{\bibfnamefont{R.}~\bibnamefont{Resta}},
  \bibinfo{journal}{Phys. Rev. Lett.} \textbf{\bibinfo{volume}{80}},
  \bibinfo{pages}{1800} (\bibinfo{year}{1998}).

\bibitem[{\citenamefont{Zak}(2000)}]{Zak:2000/PRL/1138}
\bibinfo{author}{\bibfnamefont{J.}~\bibnamefont{Zak}}, \bibinfo{journal}{Phys.
  Rev. Lett.} \textbf{\bibinfo{volume}{85}}, \bibinfo{pages}{1138}
  (\bibinfo{year}{2000}).

\bibitem[{\citenamefont{Berry}(1984)}]{Berry:1984/PRSA/45}
\bibinfo{author}{\bibfnamefont{M.~V.} \bibnamefont{Berry}},
  \bibinfo{journal}{Proc. R. Soc. A} \textbf{\bibinfo{volume}{392}},
  \bibinfo{pages}{45} (\bibinfo{year}{1984}).

\bibitem[{\citenamefont{Resta}(2000)}]{Resta:2000/JPC/R107}
\bibinfo{author}{\bibfnamefont{R.}~\bibnamefont{Resta}}, \bibinfo{journal}{J.
  Phys.: Condens. Matter} \textbf{\bibinfo{volume}{12}}, \bibinfo{pages}{R107}
  (\bibinfo{year}{2000}).

\bibitem[{\citenamefont{Bishop et~al.}(2001)\citenamefont{Bishop, Gu, and
  Kirtman}}]{Bishop:2001/JCP/7633}
\bibinfo{author}{\bibfnamefont{D.}~\bibnamefont{Bishop}},
  \bibinfo{author}{\bibfnamefont{F.~L.} \bibnamefont{Gu}}, \bibnamefont{and}
  \bibinfo{author}{\bibfnamefont{B.}~\bibnamefont{Kirtman}},
  \bibinfo{journal}{J. Chem. Phys.} \textbf{\bibinfo{volume}{114}},
  \bibinfo{pages}{7633} (\bibinfo{year}{2001}).

\bibitem[{\citenamefont{Kirtman et~al.}(2000)\citenamefont{Kirtman, Gu, and
  Bishop}}]{Kirtman:2000/JCP/1294}
\bibinfo{author}{\bibfnamefont{B.}~\bibnamefont{Kirtman}},
  \bibinfo{author}{\bibfnamefont{F.~L.} \bibnamefont{Gu}}, \bibnamefont{and}
  \bibinfo{author}{\bibfnamefont{D.~M.} \bibnamefont{Bishop}},
  \bibinfo{journal}{J. Chem. Phys.} \textbf{\bibinfo{volume}{113}},
  \bibinfo{pages}{1294} (\bibinfo{year}{2000}).

\bibitem[{\citenamefont{Springborg and
  Kirtman}(2007)}]{Springborg:2007/JCP/104107}
\bibinfo{author}{\bibfnamefont{M.}~\bibnamefont{Springborg}} \bibnamefont{and}
  \bibinfo{author}{\bibfnamefont{B.}~\bibnamefont{Kirtman}},
  \bibinfo{journal}{J. Chem. Phys.} \textbf{\bibinfo{volume}{126}},
  \bibinfo{pages}{104107} (\bibinfo{year}{2007}).

\bibitem[{\citenamefont{Sai et~al.}(2002)\citenamefont{Sai, Rabe, and
  Vanderbilt}}]{Sai:2002/PRB/104108}
\bibinfo{author}{\bibfnamefont{N.}~\bibnamefont{Sai}},
  \bibinfo{author}{\bibfnamefont{K.~M.} \bibnamefont{Rabe}}, \bibnamefont{and}
  \bibinfo{author}{\bibfnamefont{D.}~\bibnamefont{Vanderbilt}},
  \bibinfo{journal}{Phys. Rev. B} \textbf{\bibinfo{volume}{66}},
  \bibinfo{pages}{104108} (\bibinfo{year}{2002}).

\bibitem[{\citenamefont{Nunes and Gonze}(2001)}]{Nunes:2001/PRB/155107}
\bibinfo{author}{\bibfnamefont{R.~W.} \bibnamefont{Nunes}} \bibnamefont{and}
  \bibinfo{author}{\bibfnamefont{X.}~\bibnamefont{Gonze}},
  \bibinfo{journal}{Phys. Rev. B} \textbf{\bibinfo{volume}{63}},
  \bibinfo{pages}{155107} (\bibinfo{year}{2001}).

\bibitem[{\citenamefont{Baroni and Resta}(1986)}]{Baroni:1986/PRB/7017}
\bibinfo{author}{\bibfnamefont{S.}~\bibnamefont{Baroni}} \bibnamefont{and}
  \bibinfo{author}{\bibfnamefont{R.}~\bibnamefont{Resta}},
  \bibinfo{journal}{Phys. Rev. B} \textbf{\bibinfo{volume}{33}},
  \bibinfo{pages}{7017} (\bibinfo{year}{1986}).

\bibitem[{\citenamefont{Hybertsen and Louie}(1987)}]{Hybertsen:1987/PRB/5585}
\bibinfo{author}{\bibfnamefont{M.~S.} \bibnamefont{Hybertsen}}
  \bibnamefont{and} \bibinfo{author}{\bibfnamefont{S.~G.} \bibnamefont{Louie}},
  \bibinfo{journal}{Phys. Rev. B} \textbf{\bibinfo{volume}{35}},
  \bibinfo{pages}{5585} (\bibinfo{year}{1987}).

\bibitem[{\citenamefont{Jacquemin et~al.}(2003)\citenamefont{Jacquemin,
  Andr\'{e}, and Champagne}}]{Jacquemin:2003/JCP/3956}
\bibinfo{author}{\bibfnamefont{D.}~\bibnamefont{Jacquemin}},
  \bibinfo{author}{\bibfnamefont{J.-M.} \bibnamefont{Andr\'{e}}},
  \bibnamefont{and}
  \bibinfo{author}{\bibfnamefont{B.}~\bibnamefont{Champagne}},
  \bibinfo{journal}{J. Chem. Phys.} \textbf{\bibinfo{volume}{118}},
  \bibinfo{pages}{3956} (\bibinfo{year}{2003}).

\bibitem[{\citenamefont{Kudin et~al.}(2007)\citenamefont{Kudin, Car, and
  Resta}}]{Kudin:2007/JCP/234101}
\bibinfo{author}{\bibfnamefont{K.~N.} \bibnamefont{Kudin}},
  \bibinfo{author}{\bibfnamefont{R.}~\bibnamefont{Car}}, \bibnamefont{and}
  \bibinfo{author}{\bibfnamefont{R.}~\bibnamefont{Resta}}, \bibinfo{journal}{J.
  Chem. Phys.} \textbf{\bibinfo{volume}{126}}, \bibinfo{pages}{234101}
  (\bibinfo{year}{2007}).

\bibitem[{\citenamefont{Izmaylov et~al.}(2006)\citenamefont{Izmaylov, Brothers,
  and Scuseria}}]{Izmaylov:2006/JCP/224105}
\bibinfo{author}{\bibfnamefont{A.~F.} \bibnamefont{Izmaylov}},
  \bibinfo{author}{\bibfnamefont{E.~N.} \bibnamefont{Brothers}},
  \bibnamefont{and} \bibinfo{author}{\bibfnamefont{G.~E.}
  \bibnamefont{Scuseria}}, \bibinfo{journal}{J. Chem. Phys.}
  \textbf{\bibinfo{volume}{125}}, \bibinfo{pages}{224105}
  (\bibinfo{year}{2006}).

\bibitem[{\citenamefont{Gill et~al.}(1991)\citenamefont{Gill, Johnson, and
  Pople}}]{Gill:1991/IJQC/745}
\bibinfo{author}{\bibfnamefont{P.~M.~W.} \bibnamefont{Gill}},
  \bibinfo{author}{\bibfnamefont{B.~G.} \bibnamefont{Johnson}},
  \bibnamefont{and} \bibinfo{author}{\bibfnamefont{J.~A.} \bibnamefont{Pople}},
  \bibinfo{journal}{Int. J. Quantum Chem.} \textbf{\bibinfo{volume}{40}},
  \bibinfo{pages}{745} (\bibinfo{year}{1991}).

\bibitem[{GDV()}]{GDV-short}
\bibinfo{note}{Gaussian Development Version, Revision E.02, M. J. Frisch, G. W.
  Trucks, H. B. Schlegel, G. E. Scuseria, {\em et al.}, Gaussian, Inc.,
  Wallingford CT, 2004}.

\bibitem[{gdv()}]{gdv:setup}
\bibinfo{note}{SCF and geometry optimization criteria were set to ``tight'',
  which requests root-mean-squared values of $1\times10^{-8}$ and
  $1\times10^{-5}$ for density matrix change and atomic forces, and DFT
  quadrature used ``ultrafine'' grids, a pruned (99,590) grid.}

\bibitem[{\citenamefont{Guo et~al.}(1997)\citenamefont{Guo, Sirois, Proynov,
  and Salahub}}]{Book/Hadzi:1997}
\bibinfo{author}{\bibfnamefont{H.}~\bibnamefont{Guo}},
  \bibinfo{author}{\bibfnamefont{S.}~\bibnamefont{Sirois}},
  \bibinfo{author}{\bibfnamefont{E.~I.} \bibnamefont{Proynov}},
  \bibnamefont{and} \bibinfo{author}{\bibfnamefont{D.~R.}
  \bibnamefont{Salahub}}, \emph{\bibinfo{title}{Theoretical Treatment of
  Hydrogen Bonding}} (\bibinfo{publisher}{Wiley},
  \bibinfo{address}{Chichester}, \bibinfo{year}{1997}).

\bibitem[{\citenamefont{Perdew et~al.}(1996)\citenamefont{Perdew, Burke, and
  Ernzerhof}}]{Perdew:1996/PRL/3865}
\bibinfo{author}{\bibfnamefont{J.~P.} \bibnamefont{Perdew}},
  \bibinfo{author}{\bibfnamefont{K.}~\bibnamefont{Burke}}, \bibnamefont{and}
  \bibinfo{author}{\bibfnamefont{M.}~\bibnamefont{Ernzerhof}},
  \bibinfo{journal}{Phys. Rev. Lett.} \textbf{\bibinfo{volume}{77}},
  \bibinfo{pages}{3865} (\bibinfo{year}{1996}).

\bibitem[{\citenamefont{Burroughes et~al.}(1990)\citenamefont{Burroughes,
  Bradley, Brown, Marks, Mackay, Friend, Burns, and
  Holmes}}]{Burroughes:1990/N/539}
\bibinfo{author}{\bibfnamefont{J.~H.} \bibnamefont{Burroughes}},
  \bibinfo{author}{\bibfnamefont{D.~D.~C.} \bibnamefont{Bradley}},
  \bibinfo{author}{\bibfnamefont{A.~R.} \bibnamefont{Brown}},
  \bibinfo{author}{\bibfnamefont{R.~N.} \bibnamefont{Marks}},
  \bibinfo{author}{\bibfnamefont{K.}~\bibnamefont{Mackay}},
  \bibinfo{author}{\bibfnamefont{R.~H.} \bibnamefont{Friend}},
  \bibinfo{author}{\bibfnamefont{P.~L.} \bibnamefont{Burns}}, \bibnamefont{and}
  \bibinfo{author}{\bibfnamefont{A.~B.} \bibnamefont{Holmes}},
  \bibinfo{journal}{Nature} \textbf{\bibinfo{volume}{347}},
  \bibinfo{pages}{539} (\bibinfo{year}{1990}).

\bibitem[{\citenamefont{Tian et~al.}(1991)\citenamefont{Tian, Zerbi, Schenk,
  and Mullen}}]{Tian:1991/JCP/3191}
\bibinfo{author}{\bibfnamefont{B.}~\bibnamefont{Tian}},
  \bibinfo{author}{\bibfnamefont{G.}~\bibnamefont{Zerbi}},
  \bibinfo{author}{\bibfnamefont{R.}~\bibnamefont{Schenk}}, \bibnamefont{and}
  \bibinfo{author}{\bibfnamefont{K.}~\bibnamefont{Mullen}},
  \bibinfo{journal}{J. Chem. Phys.} \textbf{\bibinfo{volume}{95}},
  \bibinfo{pages}{3191} (\bibinfo{year}{1991}).

\bibitem[{\citenamefont{Rakovic
  et~al.}(1990)\citenamefont{Rakovic,
  Kostic, Gribov, and
  Davidova}}]{Rakovic:1990/PRB/10744}
\bibinfo{author}{\bibfnamefont{D.}~\bibnamefont{Rakovic}},
  \bibinfo{author}{\bibfnamefont{R.}~\bibnamefont{Kostic}},
  \bibinfo{author}{\bibfnamefont{L.~A.} \bibnamefont{Gribov}},
  \bibnamefont{and} \bibinfo{author}{\bibfnamefont{I.~E.}
  \bibnamefont{Davidova}}, \bibinfo{journal}{Phys. Rev. B}
  \textbf{\bibinfo{volume}{41}}, \bibinfo{pages}{10744} (\bibinfo{year}{1990}).

\bibitem[{\citenamefont{Vosko et~al.}(1980)\citenamefont{Vosko, Wilk, and
  Nusair}}]{Vosko:1980/CJP/1200}
\bibinfo{author}{\bibfnamefont{S.~H.} \bibnamefont{Vosko}},
  \bibinfo{author}{\bibfnamefont{L.}~\bibnamefont{Wilk}}, \bibnamefont{and}
  \bibinfo{author}{\bibfnamefont{M.}~\bibnamefont{Nusair}},
  \bibinfo{journal}{Can. J. Phys.} \textbf{\bibinfo{volume}{58}},
  \bibinfo{pages}{1200} (\bibinfo{year}{1980}).

\bibitem[{\citenamefont{Tao et~al.}(2003)\citenamefont{Tao, Perdew, Staroverov,
  and Scuseria}}]{Tao:2003/PRL/146401}
\bibinfo{author}{\bibfnamefont{J.}~\bibnamefont{Tao}},
  \bibinfo{author}{\bibfnamefont{J.~P.} \bibnamefont{Perdew}},
  \bibinfo{author}{\bibfnamefont{V.~N.} \bibnamefont{Staroverov}},
  \bibnamefont{and} \bibinfo{author}{\bibfnamefont{G.~E.}
  \bibnamefont{Scuseria}}, \bibinfo{journal}{Phys. Rev. Lett.}
  \textbf{\bibinfo{volume}{91}}, \bibinfo{pages}{146401}
  (\bibinfo{year}{2003}).

\bibitem[{\citenamefont{Vydrov and Scuseria}(2006)}]{Vydrov:2006/JCP/234109}
\bibinfo{author}{\bibfnamefont{O.~A.} \bibnamefont{Vydrov}} \bibnamefont{and}
  \bibinfo{author}{\bibfnamefont{G.~E.} \bibnamefont{Scuseria}},
  \bibinfo{journal}{J. Chem. Phys.} \textbf{\bibinfo{volume}{125}},
  \bibinfo{pages}{234109} (\bibinfo{year}{2006}).

\bibitem[{\citenamefont{Staroverov et~al.}(2003)\citenamefont{Staroverov,
  Scuseria, Tao, and Perdew}}]{Staroverov:2003/JCP/12129}
\bibinfo{author}{\bibfnamefont{V.~N.} \bibnamefont{Staroverov}},
  \bibinfo{author}{\bibfnamefont{G.~E.} \bibnamefont{Scuseria}},
  \bibinfo{author}{\bibfnamefont{J.}~\bibnamefont{Tao}}, \bibnamefont{and}
  \bibinfo{author}{\bibfnamefont{J.~P.} \bibnamefont{Perdew}},
  \bibinfo{journal}{J. Chem. Phys.} \textbf{\bibinfo{volume}{119}},
  \bibinfo{pages}{12129} (\bibinfo{year}{2003}).

\bibitem[{\citenamefont{Champagne et~al.}(1998)\citenamefont{Champagne,
  Perpete, van Gisbergen, Baerends, Snijders, Soubra-Ghaoui, Robins, and
  Kirtman}}]{Champagne:1998/JCP/10489}
\bibinfo{author}{\bibfnamefont{B.}~\bibnamefont{Champagne}},
  \bibinfo{author}{\bibfnamefont{E.~A.} \bibnamefont{Perpete}},
  \bibinfo{author}{\bibfnamefont{S.~J.~A.} \bibnamefont{van Gisbergen}},
  \bibinfo{author}{\bibfnamefont{E.-J.} \bibnamefont{Baerends}},
  \bibinfo{author}{\bibfnamefont{J.~G.} \bibnamefont{Snijders}},
  \bibinfo{author}{\bibfnamefont{C.}~\bibnamefont{Soubra-Ghaoui}},
  \bibinfo{author}{\bibfnamefont{K.~A.} \bibnamefont{Robins}},
  \bibnamefont{and} \bibinfo{author}{\bibfnamefont{B.}~\bibnamefont{Kirtman}},
  \bibinfo{journal}{J. Chem. Phys.} \textbf{\bibinfo{volume}{109}},
  \bibinfo{pages}{10489} (\bibinfo{year}{1998}).

\bibitem[{\citenamefont{Jacquemin et~al.}(2007)\citenamefont{Jacquemin,
  Perp\`{e}te, Scalmani, Frisch, Kobayashi, and
  Adamo}}]{Jacquemin:2007/JCP/144105}
\bibinfo{author}{\bibfnamefont{D.}~\bibnamefont{Jacquemin}},
  \bibinfo{author}{\bibfnamefont{E.~A.} \bibnamefont{Perp\`{e}te}},
  \bibinfo{author}{\bibfnamefont{G.}~\bibnamefont{Scalmani}},
  \bibinfo{author}{\bibfnamefont{M.~J.} \bibnamefont{Frisch}},
  \bibinfo{author}{\bibfnamefont{R.}~\bibnamefont{Kobayashi}},
  \bibnamefont{and} \bibinfo{author}{\bibfnamefont{C.}~\bibnamefont{Adamo}},
  \bibinfo{journal}{J. Chem. Phys.} \textbf{\bibinfo{volume}{126}},
  \bibinfo{pages}{144105} (\bibinfo{year}{2007}).

\end{thebibliography}

\end{document}